# Casimir Torque for a Perfectly Conducting Wedge: A Canonical Quantum Field Theoretical Approach

**H. Razmi** [1,2⊗] and **S. M. Modarresi** [2*]

1. Department of Physics, The University of Qom, Qom, I. R. Iran.
2. Department of Physics, International Imam Khomeini University, Qazvin, I. R. Iran.

## Abstract

The torque density per unit height exerted on a perfectly conducting wedge due to the quantum vacuum fluctuations (the Casimir torque) is obtained. A canonical quantum field theoretical approach with the method of calculating vacuum-to-vacuum propagator (Green function) [1] is used.

⊗ razmi@qom.ac.ir & razmiha@hotmail.com
* mo_modarresi@yahoo.com

## Introduction

More than a half-century after introducing the Casimir effect [1], it is easy to find many works, including experiments, on the Casimir force for different geometrical shapes and media by different methods of calculation [1-2]. There is a few works on the geometry of wedge shape [3-6] of them few ones deal with the Green function method [3-5].

The wedge geometry maybe considered as a natural and one of the first generalizations of the simple parallel plates geometry originally introduced by Casimir [7] in which the Casimir torque is experienced. This attractive geometrical system because of its cylindrical symmetry may have useful applications and counterparts in other parts of physics such as the cosmic string space-time [8].

As a first step toward realizing the imagination of making "vacuum machine (vacuum engineering)" [9] and microelectromechanical devices [10-12], the wedge geometry should be of special interest. This is because in any machine, rotation(s) and torque(s) are present.
Here, at first, we find the propagator (Green function) for a real scalar massless field with the geometry of the perfectly conducting wedge with an angle β; then, we calculate the vacuum-to-vacuum expectation value of the time-ordered product of the scalar field operators at two space-time points. Finally, the energy-momentum tensor and the desired Casimir torque are found. We should also mention that although this work is in a near and similar form and result to the references [3-5], where all of them deal with dyadic and explicit forms of the electric and magnetic fields; here, we work with the canonical quantum field theory of real scalar massless fields with an approach based on and similar to the method used in [1].

## The Green Function Method

The Green function method is a beautiful and powerful method based on some concepts and calculations from quantum field theory. In this method, to find the Casimir force (torque), we find from one hand, a relation between the Green function of a special (here a wedge) geometry and the vacuum expectation value for the fields operators; and on the other hand, the relation between the vacuum expectation value for the fields' operators and the energy-momentum tensor is found. As explained above, for the electromagnetic fields with two real degrees of freedom, we can simply work with the real scalar massless fields that satisfy the following Klein-Gordon equation [13]

$$(\nabla^2 - \frac{1}{c^2}\frac{\partial^2}{\partial t^2})\phi = 0 \qquad (1)$$

The corresponding equation for the time-dependent Green function (propagator) is

$$\left(-\nabla^2 + \frac{1}{c^2}\frac{\partial^2}{\partial t^2}\right)G(\vec{x},t,\vec{x}',t') = -\delta(\vec{x}-\vec{x}')\delta(t-t') \qquad (2)$$

From quantum theory of fields [13], we know that the fields' propagator, with the condition of complex frequency rotation, is

$$G(\vec{x},t,\vec{x}',t') = \frac{-i}{\hbar c}\langle 0|T\phi(\vec{x},t)\phi(\vec{x}',t')|0\rangle \tag{3}$$

in which $\langle 0|T\phi(\vec{x},t)\phi(\vec{x}',t')|0\rangle$ is the time-ordered product of fields to keep the local (causal) property of theory.

The corresponding energy-momentum density tensor for the real scalar massless fields is

$$T^{\mu\nu} = \partial^\mu\phi\partial^\nu\phi - 1/2 g^{\mu\nu}\partial^\lambda\phi\partial_\lambda\phi \tag{4}$$

With a little algebra and by means of this fact that we can simply add a total derivative to the Lagrangian (and also energy-momentum) density and by applying the equation of motion (1), we arrive at the result (the metric signature used here is $g_{00}= -g_{11}= -g_{22}= -g_{33}=1$)

$$T^{ii} = \frac{1}{2}\frac{\partial\phi}{\partial x_i}\frac{\partial\phi}{\partial x_i} \tag{5}$$

Using (5) and (3), we find the important relation

$$\langle T^{ii}\rangle = \frac{1}{2}\frac{\partial}{\partial x_i}\frac{\partial}{\partial x_i'}\langle 0|\phi(x)\phi(x')|0\rangle\big|_{x'\to x} = \frac{\hbar c i}{2}\frac{\partial}{\partial x_i}\frac{\partial}{\partial x_i'}G(x,x') \tag{6}$$

where the complex frequency rotation should be considered. Note that $i$ runs through 1 to 3 and $x = (\vec{x},t), x'=(\vec{x}',t')$.

**Time dependent Green function (propagator) for a perfectly conducting wedge with an opening angle β**

Assuming the boundary condition is of Dirichlet type, the desired Green function in terms of the complete set of (sin, cos) eigenfunctions is

$$G(\vec{x},t,\vec{x}',t') = \frac{2}{\beta}\sum_{m=1}^{\infty}\int \frac{d\omega dk}{(2\pi)^2}e^{-ik(z-z')}e^{-i\omega(t-t')}R_m(\rho,\rho')\sin(\frac{m\pi\varphi}{\beta})\sin(\frac{m\pi\varphi'}{\beta}) \tag{7}$$

in which $R_m(\rho,\rho')$ satisfies

$$\frac{d^2 R_m}{d\rho^2} + \frac{1}{\rho}\frac{dR_m}{d\rho} + (\lambda^2 - (\frac{m\pi}{\beta\rho})^2)R_m = \frac{1}{c\rho}\delta(\rho-\rho') \tag{8}$$

where $\lambda = \sqrt{\dfrac{\omega^2}{c^2} - k^2}$ .

With applying the boundary conditions and Wronskian relation for Bessel functions [14], one finds

$$G(\vec{x},t,\vec{x}',t') = \frac{1}{2c^2\beta}\sum_{m=1}^{\infty}\sin(\frac{m\pi\varphi}{\beta})\sin(\frac{m\pi\varphi'}{\beta})\int d\omega dk [e^{-ik(z-z')}e^{-i\omega(t-t')}J_\nu(\lambda\rho_<)N_\nu(\lambda\rho_>)] \quad (9)$$

where $J_\nu, N_\nu$ functions are Bessel and Neumann functions respectively and $\rho_>(\rho_<)$ is the greater (smaller) one among $\rho$ and $\rho'$. Note that $\nu = \dfrac{m\pi}{\beta}$ .

In (9), the terms containing $J_\nu(\lambda\rho_>)$ and $N_\nu(\lambda\rho_<)$ have been omitted because these functions diverge for $\rho, \rho' \to \infty, 0$ respectively.

**The Casimir torque for a perfectly conducting wedge with an opening angle β**

To calculate this vacuum torque, we should at first find the vacuum expectation value for the $\varphi\varphi$ component of the energy-momentum density tensor at $\varphi = 0, \beta$ which is (by means of (6))

$$\left\langle T^{\varphi\varphi}\big|_{\varphi=0,\beta}\right\rangle = \lim \frac{\hbar c i}{2\rho^2}\frac{\partial}{\partial\varphi}\frac{\partial}{\partial\varphi'}G(\vec{x},t,\vec{x}',t')\bigg|_{\substack{\vec{x}'\to\vec{x},t'\to t\\ \varphi'\to\varphi\to 0,\beta}} = \sum_{m=1}^{\infty}\frac{\hbar c i}{4\pi c\beta\rho^2}\left(\frac{m^2\pi^2}{\beta^2}\right)$$

$$\int d\omega dk \cos^2(\frac{m\pi\varphi}{\beta})\bigg|_{\varphi=0,\beta}\left[J_{\frac{m\pi}{\beta}}(\lambda\rho)N_{\frac{m\pi}{\beta}}(\lambda\rho)\right] \quad (10)$$

By considering the complex frequency rotation, the differential equation (8) changes to the modified Bessel equation and the above result becomes

$$\left\langle T^{\varphi\varphi}\big|_{\varphi=0,\beta}\right\rangle = \sum_{m=1}^{\infty}\frac{-\pi\hbar i}{4\pi\beta\rho^2}\left(\frac{m^2\pi^2}{\beta^2}\right)\int id\omega dk[I_{\frac{m\pi}{\beta}}(\lambda'\rho)K_{\frac{m\pi}{\beta}}(\lambda'\rho)] \quad (11)$$

where $\lambda' = \dfrac{1}{c}\sqrt{\omega^2 c^2 + k^2}$ and $I_\nu, K_\nu$ are the modified Bessel functions of the first and second kind respectively.

By changing the rectangular to polar coordinates $\int d\omega dk = \dfrac{1}{c}\int_0^{2\pi}d\theta\int_0^\infty \lambda' d\lambda'$ , we find

$$\left\langle T^{\varphi\varphi}\big|_{\varphi=0,\beta}\right\rangle = \sum_{m=1}^{\infty}\frac{\pi\hbar}{2\beta c\rho^2}\left(\frac{m^2\pi^2}{\beta^2}\right)\int_0^\infty \lambda'[I_{\frac{m\pi}{\beta}}(\lambda'\rho)K_{\frac{m\pi}{\beta}}(\lambda'\rho)]d\lambda' \quad (12)$$

Using the integral formula [15]

$$\int_0^\infty \lambda' d\lambda' [I_{\frac{m\pi}{\beta}}(\lambda'\rho) K_{\frac{m\pi}{\beta}}(\lambda'\rho)] = \lim_{\xi \to 1} \frac{\xi^{(\frac{m\pi}{\beta})}}{\rho(1-\xi^2)} \quad (13)$$

results in

$$\left\langle T^{\varphi\varphi}\big|_{\varphi=0,\beta} \right\rangle = \lim_{\xi \to 1} \left( \sum_{m=1}^{\infty} \frac{\hbar c}{2\beta\pi\rho^4} \left( \frac{m^2\pi^2}{\beta^2} \right) \right) \frac{\xi^{(\frac{m\pi}{\beta})}}{(1-\xi^2)} \quad (14)$$

To renormalize the above value for $T^{\varphi\varphi}$, it is enough to subtract $T^{\varphi\varphi}\big|_{\beta=\pi}$ from it. This is because $T^{\varphi\varphi}\big|_{\beta=\pi}$ is precisely the singular value that $T^{\varphi\varphi}$ would have if the boundary were absent. Thus

$$\left\langle T^{\varphi\varphi}\big|_{ren.} \right\rangle = \lim_{\xi \to 1} \left[ \left( \sum_{m=1}^{\infty} \frac{\hbar c}{2\beta\pi\rho^4} \left( \frac{m^2\pi^2}{\beta^2} \right) \frac{\xi^{(\frac{m\pi}{\beta})}}{(1-\xi^2)} \right) - \left( \sum_{m=1}^{\infty} \frac{\hbar c m^2}{2\pi^2 \rho^4} \frac{\xi^m}{(1-\xi^2)} \right) \right] \quad (15)$$

With some algebra, we arrive at the result

$$\left\langle T^{\varphi\varphi}\big|_{ren.} \right\rangle = \frac{-\hbar c}{480\pi^2 \rho^4} \left( \frac{\pi^4}{\beta^4} - 1 \right) \quad (16)$$

Now, we can find the desired torque density (per unit height) $N$

$$N = \frac{-1}{\rho} \frac{\partial}{\partial \beta} \left\langle T^{\phi\phi}\big|_{ren.} \right\rangle = \frac{-\pi^2 \hbar c}{120 \rho^5 \beta^5} \quad (17)$$

where $\beta = \pi$ is excluded because $\left\langle T^{\varphi\varphi}\big|_{ren.} \right\rangle$ is automatically zero for this value.

**Interpretation and asymptotic behavior**

The vacuum fluctuations due to the wedge boundary result in the torque density per unit height (17) that shows there is an attraction between the two plates of the wedge. The $\rho^{-4}$ dependence in $\left\langle T^{\varphi\varphi}\big|_{ren.} \right\rangle$ is a general property of the boundary effects in quantum field theory [16].

Meanwhile, in the limit of $\beta \to 0$ and $\rho \to \infty$ such that $\rho\beta = consant = d$, it can be shown that the energy density $T^{00}$ (just as $T^{\varphi\varphi}$ here) approaches $(-\frac{\pi^2 \hbar c}{1440 d^4})$ which is the same result as that of the two parallel plates separated by a distance *d* originally introduced by Casimir [7].